\documentclass[twocolumn,prl,showpacs,showkeys,preprintnumbers,amsmath,amssymb,floatfix]{revtex4}

\usepackage{graphicx}% Include figure files
\usepackage{dcolumn}% Align table columns on decimal point
\usepackage{bm}% bold math

\begin{document}

\preprint{}

\title{QCD Sum Rule Study of the Masses of Light Tetraquark Scalar Mesons}% Force line breaks with \\

% $s$-wave

\author{Hua-Xing Chen$^{1,2}$}
\email{hxchen@rcnp.osaka-u.ac.jp}
\author{Atsushi Hosaka$^1$}
\email{hosaka@rcnp.osaka-u.ac.jp}
\author{Shi-Lin Zhu$^2$}
\email{zhusl@th.phy.pku.edu.cn}
\address{$^1$Research Center for Nuclear Physics, Osaka
University, Ibaraki 567--0047, Japan \\$^2$Department of Physics,
Peking University, Beijing 100871, China }

\date{\today}% It is always \today, today,
         %  but any date may be explicitly specified

\begin{abstract}
We study the low-lying scalar mesons of light $u, d, s$ flavors in
the QCD sum rule. Having all possible combinations of tetraquark
currents in the local form, QCD sum rule analysis has been carefully
performed. We found that using the appropriate tetraquark currents,
the masses of $\sigma$, $\kappa$, $f_0$ and $a_0$ mesons appear in
the region of 0.6 -- 1 GeV with the expected ordering. The results
are compared with that of the conventional $\bar q q$ currents,
where the masses are considerably larger.
\end{abstract}

% PACS, the Physics and Astronomy
                 % Classification Scheme.

% 14.20.-c : Baryons (including antiparticles)
% 11.30.Rd : Chiral symmetries
% 11.30.Hv : Flavor symmetries

%Use showkeys class option if keyword
                  %display desired

% low energy theorem?

\maketitle

% introduction

The light scalar mesons have been subject to intensive discussions
for many years~\cite{Amsler:2004ps}. The expected members are
$\sigma(600)$, $\kappa(800)$, $f_0(980)$ and $a_0(980)$ of flavor
SU(3) nonet. The existence of $\sigma(600)$ which is denoted as
$f_0(600)$ in the Particle Data Group has been confirmed also by a
model independent theoretical analysis~\cite{Caprini:2005zr}.

Yet, their nature is not fully understood~\cite{scalar,Yao:2006px}.
Because they have the same spin and parity as the vacuum, $J^P =
0^+$, they may couple to many different modes. In the conventional
quark model, they are $^3P_0$ state of $\bar q q$. Their masses are,
however, expected to be larger than 1 GeV due to the $p$-wave
orbital excitation. Furthermore, the mass ordering in a naive quark
mass counting of $m_u \sim m_d < m_s$ implies $m_\sigma \sim m_{a_0}
< m_\kappa < m_{f_0}$. In chiral models, they are regarded as chiral
partners of the Nambu-Goldstone bosons ($\pi, K, \eta,
\eta^\prime)$~\cite{Hatsuda:1994pi}. Due to the collective nature,
their masses are expected to be lower than those of the quark model.
In chiral perturbation theories they are also described as
resonances of meson-meson scattering, whose quark content is
dominated by $(\bar q q)^2$~\cite{Oller:1997ti}. Recent discussions
on the scalar mesons are then largely motivated by its tetraquark
components.

Tetraquark structure of the scalar mesons was proposed long ago by
Jaffe with an assumption of strong diquark
correlations~\cite{Jaffe:1976ig}. Due to the strong attraction in
the scalar diquark channel, their masses are expected to be around
0.6 -- 1 GeV with the ordering  of $m_\sigma <  m_\kappa < m_{f_0,
a_0}$, consistent with the experimental observation
~\cite{scalar,Yao:2006px,experiment}. The form of $q \bar q$-$q \bar
q$ was also proposed~\cite{Weinstein:1982gc}. Recent lattice study
also showed an indication of the tetraquark for
$f_0$~\cite{Suganuma:2005ds}.

If such tetraquarks survive, they may be added to members of exotic
multiquark states. The subject of the multiquarks  is important,
providing a new opportunity to study colored dynamics which has not
been reached by conventional hadrons~\cite{Jaffe:2004ph}. The
problem is also related to the origin of hadron mass as we will
briefly discuss here.

In this letter, we would like to report the results of a systematic
study of the masses of the tetraquark scalar mesons in the QCD sum
rule. In the QCD sum rule one extracts hadron properties from
two-point correlation functions computed by the operator product
expansion (OPE) of QCD. The non-perturbative effects are  then
incorporated by vacuum condensates of QCD operators. By comparing
the theoretical correlation functions to phenomenological ones, one
can determine physical quantities such as masses and coupling
constants~\cite{Shifman:1978bx,Reinders:1984sr}.

QCD sum rule analyses become subtle for hadrons containing more
quarks such as tetraquarks and pentaquarks. Due to high
dimensionality of the interpolating field, one needs to calculate
many terms in OPE with high dimension. At the same time, it becomes
difficult to find a good Borel window with keeping the convergence
of OPE. Another point we would like to address here is the proper
choice of hadronic currents (interpolating fields). When the OPE has
to be in any way truncated up to certain terms, unless the current
is suitably chosen, the resulting OPE and the sum rule may not work.
In general, there are several independent currents for a given
hadron state. The optimal current can be searched by making their
linear combinations, as has been tested recently for the exotic
tetraquark state~\cite{Chen:2006hy}.

Let us construct tetraquark currents of $J^{PC} = 0^{++}$, by
establishing the number of independent currents. Following the
method in our previous work~\cite{Chen:2006hy}, we adopt the diquark
construction, where the diquark and antidiquark have the same color,
spin and orbital symmetries. Therefore, they must have the same
flavor symmetry, which is either symmetric ($\mathbf{6_f}(qq)
\otimes \mathbf{\bar 6_f}(\bar q \bar q)$) or antisymmetric
($\mathbf{\bar 3_f}(qq) \otimes \mathbf{3_f}(\bar q \bar q)$). Then
we assume the ideal mixing in which only isospin symmetry is
respected and the currents are classified by the number of strange
quarks. Hence, denoting light $u, d$ quarks by $q$, $\sigma$
currents are constructed as $qq \bar q \bar q$, $\kappa$ currents by
$qs \bar q \bar q$ and $f_0$ and $a_0$ currents by $qs \bar q \bar
s$.

Using the antisymmetric combination for diquark flavor structure, we
arrive at the following five independent currents
\begin{eqnarray}
\nonumber\label{define_udud_current} S^\sigma_3 &=& (u_a^T C
\gamma_5 d_b)(\bar{u}_a \gamma_5 C \bar{d}_b^T - \bar{u}_b \gamma_5
C \bar{d}_a^T)\, ,
\\ \nonumber
V^\sigma_3 &=& (u_a^T C \gamma_{\mu} \gamma_5 d_b)(\bar{u}_a
\gamma^{\mu}\gamma_5 C \bar{d}_b^T - \bar{u}_b \gamma^{\mu}\gamma_5
C \bar{d}_a^T)\, ,
\\
T^\sigma_6 &=& (u_a^T C \sigma_{\mu\nu} d_b)(\bar{u}_a
\sigma^{\mu\nu} C \bar{d}_b^T + \bar{u}_b \sigma^{\mu\nu} C
\bar{d}_a^T)\, ,
\\ \nonumber
A^\sigma_6 &=& (u_a^T C \gamma_{\mu} d_b)(\bar{u}_a \gamma^{\mu} C
\bar{d}_b^T + \bar{u}_b \gamma^{\mu} C \bar{d}_a^T)\, ,
\\ \nonumber
P^\sigma_3 &=& (u_a^T C d_b)(\bar{u}_a C \bar{d}_b^T - \bar{u}_b C
\bar{d}_a^T)\, ,
\end{eqnarray}
where the sum over repeated indices ($\mu$, $\nu, \cdots$ for Dirac,
and $a, b, \cdots$ for color indices) is taken. Either plus or minus
sign in the second parentheses ensures that the diquarks form the
antisymmetric combination in the flavor space. The currents $S$,
$V$, $T$, $A$ and $P$ are constructed by scalar, vector, tensor,
axial-vector, pseudoscalar diquark and antidiquark fields,
respectively. The subscripts $3$ and $6$ denote the color states of
the diquarks (antidiquarks) which are combined into the color
representation $\mathbf{\bar 3_c}$ and $\mathbf{6_c}$
($\mathbf{3_c}$ or $\mathbf{\bar 6_c}$), respectively. The currents
for other members are formed by the following replacements in
(\ref{define_udud_current}), $\kappa: \; (ud) (\bar u \bar d)
\to(ud) (\bar d \bar s)$, $f_0: \; (ud) (\bar u \bar d) \to(us)
(\bar u \bar s) + (u \leftrightarrow d)$, $ a_0: \; (ud) (\bar u
\bar d) \to(us) (\bar u \bar s) - (u \leftrightarrow d)$. More
details will be discussed in a separate
publication~\cite{Chen:2006pre}.

Using the tetraquark current $\eta$ which is one of the currents of
(\ref{define_udud_current}) or their linear combination, we have
computed the correlation function
\begin{equation}
\Pi_{\rm OPE}(q^2)\,\equiv\,i\int d^4x e^{iqx}
\langle0|T\eta(x){\eta^\dagger}(0)|0\rangle \, , \label{eq_pidefine}
\end{equation}
in the OPE up to dimension eight, keeping the current quark masses
$m_u$, $m_d$ and $m_s$ finite. As the primary requirement, the
spectral densities $\rho_{\rm OPE}$ must be positive definite. If
truncation of OPE is not good, it happens that they become negative
sometimes. Using the dispersion relation, it is equivalently written
as
\begin{equation}
\Pi_{\rm OPE}(q^2)\,\equiv\, \int_0^\infty ds \frac{\rho_{\rm
OPE}(s)}{s - q^2 - i \epsilon} \, ,
\end{equation}
where $\rho_{\rm OPE} = {\rm Im} \Pi_{\rm OPE} / \pi$. This is then
equated to the integral over the physical (phenomenological)
spectral density $\rho_{\rm phen}(s)$.

The phenomenological spectral density is parameterized as a sum of
one pole and continuum contributions. Assuming that the continuum
part is approximated by the one of OPE
(duality)~\cite{Reinders:1984sr},
\begin{equation}\label{eq_phen}
\rho_{\rm phen}(s) = f^2 \delta(s-M^2) + \theta(s-s_0)
\frac{1}{\pi}{\rm Im} \Pi_{\rm OPE}(s) \, ,
\end{equation}
where $M$ and $f$ are the mass and coupling constant of the physical
state under investigation. In order to extract physical quantities
efficiently by suppressing the continuum contribution, the Borel
transformation is performed. Finally we arrive at the sum rule
equation
\begin{equation}
f^2 \exp(-M^2/M_B^2) = \int_0^{s_0} ds \rho_{\rm OPE}(s)
e^{-s/M_B^2}\, ,
\end{equation}
which determines the mass and the coupling constant. The mass $M$ is
a function of the two parameters $s_0$ and $M_B$. They must be
chosen to satisfy (1) rapid convergence of OPE, (2) sufficient
amount of pole contribution and (3) weak dependence on $s_0$ and
$M_B$. These are important in order to draw reliable
conclusions~\cite{Kojo:2006bh}.

The use of the $\delta$-function in (\ref{eq_phen}) for the scalar
mesons might be subject to criticisms, firstly because the observed
scalar mesons have wide decay width. The inclusion of a finite width
in the QCD sum rule instead of using the $\delta$-function is one
option to take care of the effect of the decay width. We have
performed such analysis in the form of Gaussian and verified that
still it is possible to reproduce experimental values of
masses~\cite{Chen:2006pre}.

Secondly, the tetraquark currents is expected to couple strongly to
two meson states. We argue that such two meson contributions can be
computed separately from the short distance method of OPE. By
applying the soft-pion theorem~\cite{Dai:2006uz}, the coupling to
two meson states can be expressed by a double commutator with the
axial charge $Q_5$, $\langle 0 | \eta | \pi^a \pi^b \rangle \sim
\langle 0 | [ Q_5 , [ Q_5 , \eta ] ] | 0 \rangle$. Since one
commutator yields the factor $\bar q q$, we find $(\bar q q)^4$
altogether in the two-point correlation function. The dimension of
this term is as high as twelve, which is beyond the present study
where we compute up to dimension eight. Similarly, as shown in
Ref.~\cite{Brito:2004tv} which also uses the method of QCD sum rule,
the coupling of $s \rightarrow pp$ ($s$ stands for a scalar
tetraquark and $p$ pseudoscalar meson) is of higher order as
proportional to $(\bar q q)^2$, consistently implying that the decay
width of the $s \rightarrow pp$ is of order $\langle \bar q q
\rangle^4$.

As we will see shortly, the fact that the present QCD sum rule with
the OPE up to dimension eight will yield a stable solution indicates
that there is a significant component in the scalar meson state
which couples to the tetraquark current without going through two
mesons. In this case, we expect that the narrow resonance
approximation is reliable. The large decay width will then be
explained through the coupling to two meson states in the form of
high-dimension terms of OPE. In fact, we have performed a QCD sum
rule analysis by using a peak of finite width, and found that the
result does not change much.

We have performed the sum rule analysis using all currents and their
various linear combinations. We have found that the results for
single currents are not reliable, except for the tensor current
$T_6^\sigma$, due to either violation of positivity or insufficient
convergence of OPE. In fact, we have found good sum rule by a linear
combination of $A_6^\sigma$ and $V_3^\sigma$: $\eta^\sigma_1 =
\cos\theta A^\sigma_6 + \sin\theta V^\sigma_3\, $, where the best
choice of the mixing angle turns out to be $\cot\theta = 1 /
\sqrt{2}$. For $\kappa$, $f_0$ and $a_0$, we have also found that
similar linear combinations give better sum rules.

The calculation of OPE is tedious but straightforward. The results
up to dimension eight are
%------------------------------\simga 1----------------------------------
\begin{eqnarray} \label{Pi_sigma}
\rho^\sigma(s) &=& {s^4 \over 11520 \pi^6}  + { (6 \sqrt{2} + 7)
\langle g^2 G G \rangle s^2 \over 9216 \pi^6}
\\
&+& (m_u + m_d) \langle \bar q q \rangle \Big( { s^2 \over 36 \pi^4}
+ { (6 \sqrt{2} + 1) \langle g^2 G G \rangle \over 1152 \pi^4 } \Big
) \nonumber\\ \nonumber&+& \mathcal{O}(m_q^2)\, ,
\end{eqnarray}
%------------------------------\kappa 1-----------------------------------
\begin{eqnarray}
\rho^\kappa(s) &=& {1 \over 11520 \pi^6} s^4 - {m_s^2 \over 572
\pi^6}  s^3
\\
&+& \big ( { 6 \sqrt{2} + 7 \over 9216 \pi^6} \langle g^2 G G
\rangle + {m_s \langle \bar s s \rangle \over 72 \pi^4} \big ) s^2
\nonumber \\
&+& \big ( - {6 \sqrt{2} + 7 \over 3072 \pi^6 } m_s^2 \langle g^2 G
G \rangle + { m_s \langle \bar q \sigma G q \rangle \over 128 \pi^4
} \big ) s
\nonumber \\
&-& { m_s \langle g^2 G G \rangle \langle \bar q q \rangle \over 384
\pi^4} - { \langle \bar s s \rangle \langle \bar q \sigma G q
\rangle \over 48 \pi^2 } \nonumber \\ \nonumber &+& { \langle \bar q
q \rangle \langle \bar s \sigma G s \rangle \over 48 \pi^2 } + { 6
\sqrt{2} + 7 \over 2304 \pi^4 } m_s \langle g^2 G G \rangle \langle
\bar s s \rangle \, , \label{Pi_kappa}
\end{eqnarray}
%------------------------------\a0 f0 1-----------------------------------
\begin{eqnarray}\label{Pi_a0f0}
\rho^{f_0}(s) &=& {1 \over 11520 \pi^6} s^4 - {m_s^2 \over 288
\pi^6}  s^3
\\
&+& \big ( { 6 \sqrt{2} + 7 \over 9216 \pi^6} \langle g^2 G G
\rangle + {m_s \langle \bar s s \rangle \over 36 \pi^4} \big ) s^2
\nonumber \\
&+& \big ( - {6 \sqrt{2} + 7 \over 1536 \pi^6 } m_s^2  \langle g^2 G
G \rangle - { m_s^3 \langle \bar s s \rangle \over 6 \pi^4 } \big )
s
\nonumber \\
&-& { m_s \langle g^2 G G \rangle \langle \bar q q \rangle \over 192
\pi^4} + { 4 m_s^2 \langle \bar q q \rangle^2 \over 9 \pi^2 }
\nonumber \\ \nonumber &+& { 4 m_s^2 \langle \bar s s \rangle^2
\over 9 \pi^2 } + { 6 \sqrt{2} + 7 \over 1152 \pi^4 } m_s \langle
g^2 G G \rangle \langle \bar s s \rangle\, .
\end{eqnarray}
The OPE for $a_0$ takes the same form as for $f_0$. For $\sigma$,
terms containing $u, d$ quark masses $m_q$ are small. For instance,
the term of $m_q \langle \bar q q \rangle$ of dimension four is
about ten times smaller than the other term of $\langle g^2 G G
\rangle$. For $\kappa$, $a_0$ and $f_0$, the terms containing
strangle quark mass are important but those containing $u$ and $d$
quark masses are negligibly small.

We use the following values of
condensates~\cite{Yang:1993bp,Ioffe:2002be,Gimenez:2005nt}:
$\langle\bar qq \rangle=-(0.240 \mbox{ GeV})^3$, $\langle\bar
ss\rangle=-(0.8\pm 0.1)\times(0.240 \mbox{ GeV})^3$,$\langle
g_s^2GG\rangle =(0.48\pm 0.14) \mbox{ GeV}^4$, $ m_u = 5.3 \mbox{
MeV}$, $m_d = 9.4 \mbox{ MeV}$, $m_s(1\mbox{ GeV})=125 \pm 20 \mbox{
MeV}$, $\langle g_s\bar q\sigma G q\rangle=-M_0^2\times\langle\bar
qq\rangle$, $M_0^2=(0.8\pm0.2)\mbox{ GeV}^2$. Now let us discuss the
feasibility of our QCD sum rule.

\begin{figure}[hbt]
\begin{center}
\scalebox{0.8} {\includegraphics{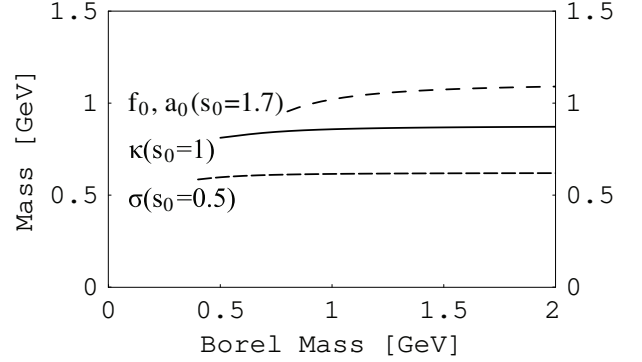}} \caption{Masses of
the $\sigma$ (short-dashed), $\kappa$ (solid), $f_0$ and $a_0$
(long-dashed) mesons calculated by the tetraquark currents as
functions of the Borel mass $M_B$, with $s_0$ (GeV$^2$) as shown in
figures.} \label{pic_tetra}
\end{center}
\end{figure}

The Borel transformed correlation functions are written as power
series of the Borel mass $M_B$. Since the Borel transformation
suppresses the contributions from $s > M_B^2$, smaller values are
preferred to suppress the continuum contributions. However, for
smaller $M_B$ convergence of the OPE becomes worse. Therefore, we
should find an optimal value of $M_B$. We have found that $M_B \sim
0.4$ GeV for $\sigma$, $0.5$ GeV for $\kappa$ and $0.8$ GeV for
$f_0$ and $a_0$, where the pole contributions reach around 50\% for
all cases, while the convergence is still sufficiently
fast~\cite{Chen:2006pre}. As $M_B$ is increased, the pole
contributions decrease, but the resulting tetraquark masses are
stable as shown in Fig.~\ref{pic_tetra}.

%---------figure Mass_MB_mixed_current

\begin{figure}[tbh]
\begin{center}
\scalebox{0.8} {\includegraphics{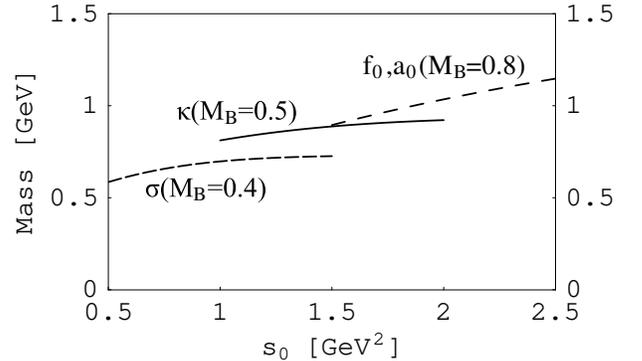}} \caption{Masses of
the $\sigma$ (short-dashed), $\kappa$ (solid), $f_0$ and $a_0$
(long-dashed) mesons calculated by the tetraquark currents as
functions of threshold value $s_0$, with $M_B$ (GeV) as shown in
figures.} \label{pic_s0}
\end{center}
\end{figure}

We have also searched the region where the tetraquark mass varies
significantly less than the change in $\sqrt{s_0}$. We have found
such regions $0.5 < s_0 ({\rm GeV}^2) < 1.5$ for $\sigma$, $1 < s_0
< 2$ for $\kappa$, and $1.5 < s_0 < 2.5$ for $f_0$ and $a_0$. In
Fig.~\ref{pic_s0}, we show $s_0$ dependence of the masses in these
regions. As we see, the mass is stable in a rather wide region of
$s_0$. The Borel mass dependence of Fig.~\ref{pic_tetra} are shown
for the minimum values of $s_0$.

%---------figure Mass_MB_mixed_current

After careful test of the sum rule for a wide range of parameter
values of $M_B$ and $s_0$, we have found reliable sum rules, with
which we find the masses $ m_\sigma = (0.6 \pm 0.1) \; {\rm GeV}$, $
m_\kappa = (0.8 \pm 0.1) \; {\rm GeV}$,  $m_{f_0,a_0} = (1 \pm 0.1)
\; {\rm GeV}\;.$ It is interesting to observe that the masses appear
roughly in the order of the number of strange quarks with roughly
equal splitting.

It would be interesting to observe from Eqs.~(\ref{Pi_sigma}) -
(\ref{Pi_a0f0}) that the mass of the $\sigma$ is dominated by the
gluon condensate, while other condensates with $m_s$ also play a
significant role for other masses. In fact in the SU(3) limit, where
all quark masses and condensates take the same values, the three
equations become identical. In particular, in the chiral limit where
all quark masses vanish, the masses of the scalar mesons are
dictated only by the gluon condensate. This property was also
observed in a recent publication~\cite{Lee:2006vk}.

Now for comparison, we have also performed the QCD sum rule analysis
using the $\bar q q$ current within the present framework, although
such works have been done
before~\cite{Reinders:1981ww,Kisslinger:1997gs,Elias:1998bq,Du:2004ki}.
We have computed the OPE up to dimension six in this case, and have
verified the previous results. Namely, the masses of the $\bar q q$
mesons are considerably heavier than the masses of the tetraquark
mesons.

%%---------figure Mass_MB_mixed_current
%\begin{figure}[tbh]
%\begin{center}
%\scalebox{0.8} {\includegraphics{mass_meson}}
%\caption{Mass of
%the $\sigma$  meson calculated by the conventional $\bar q
%q$ current  as functions of the Borel mass, $M_B$ MeV.} \label{pic_meson}
%\end{center}
%\end{figure}

In conclusion we have found that the QCD sum rule analysis with
tetraquark currents implies the masses of scalar mesons in the
region of 600 -- 1000 MeV with the ordering, $m_\sigma < m_\kappa <
m_{f_0, a_0}$, while the conventional $\bar q q$ currents are
considerably heavier. Our conclusion has become rather robust, after
we have tested all possible independent tetraquark currents and with
their linear combinations.

Our observation supports a tetraquark structure for low-lying scalar
mesons. Somewhat non-trivial is that a large part of the mass is due
to the gluon condensate rather than chiral condensate. This
observation is interesting in relation to the question of the origin
of the mass generation of hadrons~\cite{Weinberg:1969hw}. To test
the validity of the tetraquark structure, it is also important to
study decay properties, which is often sensitive to the  structure
of wave functions. Such a tetraquark structure will open an
alternative path toward the understanding exotic multiquark dynamics
which one does not experience in the conventional hadrons.

{\bf Acknowledgments:} H.X.C. is grateful to the Monkasho fellowship
for supporting his stay at Research Center for Nuclear Physics where
this work is done. A.H. is supported in part by the Grant for
Scientific Research ((C) No.16540252) from the Ministry of
Education, Culture, Science and Technology, Japan. S.L.Z. was
supported by the National Natural Science Foundation of China under
Grants 10421503 and 10625521, Ministry of Education of China,
FANEDD, Key Grant Project of Chinese Ministry of Education (NO
305001) and SRF for ROCS, SEM.

\end{document}